\documentclass[12pt]{article}
\renewcommand{\bar}[1]{\overline{#1}}
\newcommand{\longvec}[1]{\overrightarrow{\!\!#1}}
\newcommand{\VEV}[1]{\left\langle{#1}\right\rangle}

\begin{document}
\begin{flushright}
{\small SLAC--PUB--9063 \\
November 2001}
\end{flushright}

\hfill

\begin{center}
{{\bf\large Emergent Gauge Bosons}\footnote{Work supported by
Department of Energy contract DE--AC03--76SF00515.}}

\bigskip
{\it James Bjorken\\
Stanford Linear Accelerator Center\\
Menlo Park, CA 94025 \\
bjorken@slac.stanford.edu}
\medskip
\end{center}
\vfill
\begin{center}
{\bf\large Abstract}
\end{center}

The old idea that the photon is a Goldstone boson emergent from a
spontaneously broken theory of interacting fermions is revisited.
It is conjectured that the gauge-potential condensate has a vacuum
expectation value which is very large, perhaps the GUT/Planck
momentum scale, but that the magnitude of the effective potential
which generates it is very small, so small that in the limit of
vanishing cosmological constant it would vanish.  In this way, the
threat of unacceptably large observable, noncovariant residual
effects is mitigated. The linkage of these ideas to other
speculative ideas involving black holes and parametrizations of
Standard-Model coupling constants is also described. \vfill

\begin{center}
{\it Contributed to the 4th Workshop  \\
``What Comes beyond the Standard Model?" \\
 Bled, Slovenia\\
July 17--27, 2001 }\\
\end{center}

\vfill \newpage

\section{Introduction}\label{sec1}

It is a pleasure to write this note on the occasion of Holger Bech
Nielsen's 60th birthday celebration. It will deal with very
speculative matters, most of which down through the years have
benefitted by interactions with him. Holger Bech is one of few
with whom I find it fruitful to even discuss such things. Not only
has he been always receptive, but most importantly he has reacted
thoughtfully and critically, instead of the commonplace reaction
of indifference.

The subject matter discussed below is actually of common interest:
whether gauge bosons such as the photon may be regarded as
emergent degrees of freedom. The word ``emergent" is used in the
condensed matter community to describe collective modes,
appropriate for a low energy description of a quantum fluid (or
other medium) built of more fundamental degrees of freedom. I will
further specify that the emergent gauge bosons be considered as
Goldstone modes, associated with a spontaneous breakdown of
Lorentz symmetry. This is an idea I entertained long ago
\cite{ref:a}, one which clearly has a lot to do with Holger's
viewpoints as well \cite{ref:b}.

Recently I have played with the notion of relating Standard Model
parameters to gravitational parameters \cite{ref:c}, and this has
prompted me to revisit the old Goldstone-photon ideas again. What
follows will be a status report on what I have found out. In the
next section I will sketch out a modernized version of ancient
work on Goldstone photons. Section 3 contains the speculations
regarding Standard Model parameters, and how those ideas mesh with
the Goldstone-photon description. Section 4 deals with potentially
observable effects of Lorentz-covariance violation, something
which is a central problem of any approach based upon emergence.
Section 5 attempts to summarize what is done in somewhat more
general language, and to connect it with some recent work of
Holger Bech and his collaborators \cite{ref:CFN}. In Section 6 we
look at some of the future problems and opportunities presented by
this approach.

\section{The Goldstone Photon}

Shortly after the work of Nambu and Jona-Lasinio \cite{ref:d} on
the ``Goldstone pion", I explored whether their method might be
applicable to the description of the photon as a Goldstone mode. A
Fermi current-current contact interaction between fermions was
posited, and then the Nambu-Jona-Lasinio methods were imitated. A
nonvanishing condensate carrying a (timelike) vacuum current was
assumed, determined by a gap equation even more obscure than the
quadratically divergent one they considered for hadronic matter.
The properties of the current-current correlation function,
however, became constrained by the gap equation in such a way as
to yield essentially a photon propagator in temporal gauge.  The
reasoning was analogous to their arguments for a massless-pion
propagator. A challenge remained, however.  It appeared that there
were physically observable manifestations of Lorentz noncovariance
occurring in higher order effects. These were concentrated in
nonlinear, gauge noncovariant, photon self-interactions. To
suppress them, the Fermi constant in the original current-current
interaction, as well as the magnitude of the vacuum current, had
to be taken as very large. The physics of these assumptions, as
well as that of the gap equation, was left up in the air.

We now review these results, using the modern language of path
integrals and effective potentials. From the standard path
integral we introduce the vector field $A_\mu$ which will mediate
the four-fermi interaction and in fact will allow the fermion
fields to be ``integrated out". Starting with the original
Lagrangian density,
\begin{equation}
\mathcal{L} = \bar\psi(i\not\! \nabla-m)\psi + \frac{G}{2}
\left(\bar\psi\gamma_\mu\psi\right)^2 \ ,
\label{eq:a}
\end{equation} we write the path integral as
\begin{eqnarray}
W(J) &=& \int \mathcal{D}\psi\mathcal{D}\bar\psi\, \exp\ i\int
d^4x\left[\mathcal{L}-\bar\psi\gamma_\mu\psi J^\mu\right]
\\[1ex] &=& N\int \mathcal{D}\psi\mathcal{D}\bar\psi\mathcal{D}A\,
\exp\ i  \int d^4x \left[\mathcal{L}-\bar\psi\gamma_\mu\psi J^\mu
- \frac{1}{2G}\left(A_\mu-G\bar\psi\gamma_\mu\psi\right)^2\right]
\ . \nonumber \label{eq:b}
\end{eqnarray} The auxiliary Gaussian
factor is designed to eliminate the four-fermi interaction,
allowing us to integrate out the fermions. We find, using standard
techniques
\begin{eqnarray}
W(J) &=& N^\prime\int \mathcal{D}A\, \exp\ -i\int d^4x \left[
\frac{A^2}{2G}- V(A-J)\right] \nonumber \\[1ex]
&=& N^\prime \int \mathcal{D}A\, \exp\ -i\int d^4x
\left[\frac{(A+J)^2}{2G}-V(A)\right] \label{eq:c}
\end{eqnarray} with $V(A)$ the one-loop effective
potential, formally given by
\begin{eqnarray}
i \int d^4x\, V(A) &=& \ell n\,
\det(i\not\!\nabla+\not\!\!A-m)-\ell n\, \det
(i\not\!\nabla-m) \nonumber \\
&=& {\rm Tr}\  \ell n \frac{1}{i\not\!\nabla-m}\,
(i\not\!\nabla+\not\!\!A-m) \\[1ex]
&=& \int d^4x\, \bigg[{\rm tr}\, \not\!\!A(x)S_F(x,x)\nonumber
\\[1ex]
&& - \left.\frac{1}{2}\, {\rm tr}\int d^4y
\not\!\!A(x)S_F(x,y)\not\!\!A(y)S_F(y,x) + \cdots\right]\nonumber
\label{eq:d}
\end{eqnarray}
and with tr implying a Dirac trace.

The formal steps up to this point are straightforward. But the
physics to be applied is less clear. We here assume that, despite
the presence of ultraviolet cutoffs, the form of the effective
potential obeys Lorentz covariance, or at least that the
corrections thereto will not disturb the line of argument to
follow.  Thus $V$ is assumed to be a function only of the
invariant $A_\mu A^\mu \equiv A^2$.  We also assume that, in the
limit of $J_\mu$ being an infinitesimal constant timelike external
source, there is a classical condensate $A_\mu$ which remains
proportional to $J_\mu$ and nonvanishing as $J_\mu$ is turned off.
That is, spontaneous Lorentz-symmetry breaking is assumed. The
equilibrium value of $A_\mu$ will be determined in the usual way
by a saddle-point integration over the fields $A$.

We need to evaluate the important contributions to $V(A)$, all of
which are lowest order fermion closed loops. We expand in powers
of $A$, keeping for now only quadratic and quartic terms. For the
quadratic piece we make a derivative expansion. The term with no
derivatives renormalizes the explicit quadratic term in the
action, proportional to $G^{-1}$. We also clearly need a kinetic
term in the action, which is quadratic in derivatives and
quadratic in $A$. In the ancient work it was, in momentum space,
\begin{eqnarray}
\int d^4x \Delta V^{(2)}(x) &=& \frac{1}{2}\int
\frac{d^4q}{(2\pi)^4} \widetilde A_\mu(q)\, \widetilde A_\nu(q)
(q_\mu q_\nu-g_{\mu\nu}q^2) \frac{1}{e^2(q^2)}\nonumber \\[1ex]
&\approx& \frac{1}{4e^2(0)} \int d^4x\, F_{\mu\nu}(x)
F^{\mu\nu}(x)
\label{eq:e}
\end{eqnarray} with $1/e^2$ given by the
usual vacuum polarization integral
\begin{equation}
\frac{1}{e^2(0)} = \frac{1}{12\pi^2} \, \ell n
\frac{\Lambda^2}{m^2} \Rightarrow
\frac{1}{12\pi^2}\int^{\Lambda^2}_{4m^2} R(s) \frac{ds}{s} \
\label{eq:f}
\end{equation} and $\Lambda$ some momentum-space
cutoff. With this definition, the running coupling constant
\begin{equation}
\frac{1}{e^2(q^2)} \cong \frac{1}{12\pi^2} \, \ell n
\frac{\Lambda^2}{q^2} \Rightarrow \frac{1}{12\pi^2}
\int^{\Lambda^2}_{q^2} R(s)\frac{ds}{s}
\label{eq:g}
\end{equation}
vanishes as $q^2\rightarrow \Lambda^2$, corresponding to the
compositeness condition $Z_3 = 0$, which implies infinite bare
charge at the highest energy scale $\Lambda$.\footnote{In those
ancient times, there was no electroweak unification.  In what
follows we make believe, with eyes wide open, that this is still
the case.}

But we here choose to add an additional contribution $1/e_0^2$,
not present in the ancient work, to account for the fact that even
at the GUT unification scale the running gauge coupling constants
are finite.  That is, we expect that there must be additional
contributions from physics beyond the Standard Model.  We shall
return to this question in the Section 5. Putting all this
together yields the terms in the effective action which are
quadratic in $A_\mu$:
\begin{equation}
S^{(2)} = \int d^4x\, \mathcal{L}^{(2)}(x) = \int d^4x
\left[\frac{A^2}{2G}+\frac{1}{4e^2}\, F_{\mu\nu} F^{\mu\nu}\right]
\ . \label{eq:h}
\end{equation} There must be quartic, and possibly
higher, terms in $V(A)$ as well, in order to generate the familiar
Mexican hat structure of the potential and induce spontaneous
symmetry breaking. At zero momentum transfer, the conventional
contributions to the quartic fermion loop vanish by current
conservation. But, just as we assumed an extra piece in the vacuum
polarization term, we also assume that the quartic term is
nonvanishing:
\begin{equation}
S^{(4)} = - \int d^4x\frac{\lambda}{4} (A^2)^2 \ .
\label{eq:i}
\end{equation} Clearly this leads to a minimum value
of the potential at
\begin{equation}
A^2_{cl} \equiv M^2 = \frac{1}{G\lambda} \ .
\label{eq:j}
\end{equation}
We may now do the path integral over the field $A_0$, leaving the
classical term in the exponent. In this short note we will drop
the determinantal prefactor, because it corresponds to photon
closed loop contributions beyond the scope of these
considerations. One can eventually do better in this regard.

What is left is an effective action similar to the effective
action of chiral perturbation theory, to be considered in tree
approximation.  In the case of pions, there are many ways to
express the effective action, depending upon how one parametrizes
the fields. This is also true in this case, and an especially
convenient way is to eliminate the time component $A_0$ in terms
of the space components.
\begin{equation}
A_0 = \sqrt{M^2+\longvec A^2} = M + \frac{\longvec A^2}{2M} +
\cdots \ .
\label{eq:k}
\end{equation}
 This is analogous to the elimination of the sigma component of
 the field in terms of the pions in the
effective action via
\begin{equation}
\sigma = \sqrt{f^2 - \longvec \pi^2} = f -
\frac{\longvec\pi^2}{2f} + \cdots \ . \label{eq:l}
\end{equation}
Such a choice provides a very efficient way of deriving the
structure of pion-pion scattering at second order in momentum.

In our case the elimination of $A_0$ in terms of the space
components leads to the effective Lagrangian
\begin{equation}
\mathcal{L} = \frac{1}{2e^2} [E^2-B^2] - V(A_{cl}) =
\frac{1}{2e^2} \left[\left(\dot{\longvec A}+\longvec\nabla
A_0\right)^2-\left(\longvec\nabla\times \longvec A\right)^2\right]
- V(A_{cl}) \label{eq:m}
\end{equation}
with $A_0$ expressed in terms of $\longvec A$ by Eq. (\ref{eq:k}).
Note that in tree approximation, to which we have limited ourself,
there is no dependence left on the nature of the potential $V$.
The complications have been transferred into the kinetic energy
portion of the Lagrangian. In the approximation of a quadratic
action, we reduce to free fields in temporal gauge. When going
beyond that approximation, there is a nonlinear term in the
effective Lagrangian  containing the time derivative of $\longvec
A$. Such correction terms are small but serious, since they appear
to violate Lorentz covariance and gauge invariance. We will return
to this in Section 4, after consideration of the orders of
magnitude of the terms we have introduced.

As we already mentioned, in the ancient work it was necessary to
assume that the vacuum value of $A_\mu $was very large
\begin{equation}
 \VEV{A_\mu} \equiv \eta_\mu M
\label{eq:n}
\end{equation}
with $\eta_\mu$ a unit timelike vector and $M$ very large. We may
naturally implement this by assuming
\begin{equation}
\lambda = \frac{\mu}{M} \label{eq:o}
\end{equation} with $\mu$ a
very small mass. Then the overall effective potential is
\begin{equation}
V_{\rm eff} \equiv  \frac{-A^2}{2G} + V(A) = \frac{-A^2}{2G} +
\frac{\mu}{4M} A^4 \equiv \frac{\mu M^3}{4}
\left[-2\left(\frac{A^2}{M^2}\right) +
\left(\frac{A^2}{M^2}\right)^2\right]
\label{eq:p}
\end{equation}
where we identify
\begin{equation}
G \equiv \frac{1}{\mu M} \ .
\label{eq:q}
\end{equation} The
general form of this overall effective potential
\begin{equation}
V_{\rm eff} = \mu M^3 F\left(\frac{A^2}{M^2}\right)
\label{eq:r}
\end{equation} may well be preserved even in the
presence of radiatively induced terms of higher order, such as
~$A^6$, $A^8$, {\em etc}. In the next section the parameter $\mu$
will be identified with the cosmological-constant scale and $M$
with the GUT or Planck scale. The small expansion parameter in the
ancient work, characterizing the importance of noncovariant
physical effects, was associated with a hypothetical
momentum-space cutoff $\Lambda$
\begin{equation}
\VEV{A_\mu} = \eta_\mu M \sim G\Lambda^3 = \frac{\Lambda^3}{M\mu}
\ . \label{eq:s}
\end{equation} Higher order effects were
characterized by powers of the dimensionless parameter
\begin{equation}
\frac{1}{G\Lambda^2} = \left(\frac{\mu}{M}\right)^{1/3}
\label{eq:t}
\end{equation} which evidently is small, perhaps small
enough to escape experimental constraints. However, there is
unfinished business in relating the scheme presented in the
ancient work to the language in this note, where the expansion is
in integer powers of $\mu/M \sim 10^{-30}$.  Some of this
unfinished business appears to involve real physics, not mere
formalism.

\section{Standard Model Parameters}

For a long time I have enjoyed contemplating the 20-parameter
Standard Model in the formal, ``gaugeless" limit of vanishing
gauge coupling constants \cite{ref:f}. Originally the motivation
was to exhibit that, unlike the claims made all too often and too
sloppily, it is not true that all known Standard Model
interactions are gauge interactions. The proof is to look at the
Standard Model in the gaugeless limit.  The electron becomes
unstable, decaying into a neutrino and massless longitudinal $W$
(the Goldstone mode, which does not decouple in the limit).
Evidently there has to be an extra, nongauge force present for
this to happen, and in the standard picture it comes from the
Higgs sector.

Now I think we all believe that the formal gaugeless limit is
artificial, and that in a better theory the other parameters will
not stay the same while gauge couplings are varied. This belief
became motivation for trying to guess how they might change. And
without going into details, I have found a scenario which relates
all the major Standard Model parameters to the Planck/GUT scale
$M$ and the cosmological constant scale $\mu$ ($\mu^4$ is the
usual cosmological constant) \cite{ref:c}. The first guess is that
\begin{equation}
\frac{1}{g^2} \sim \frac{1}{h^2} \sim \frac{1}{\lambda} \sim
\frac{1}{4\pi^2}\log \frac{M^2}{\mu^2}
\label{eq:u}
\end{equation}
where $g$ is the gauge coupling constant, $h$ the top quark
coupling constant, and $\lambda$ is the Higgs quartic coupling,
all evaluated at the GUT scale. The large logarithm explains why
these couplings are small.

The second guess is that the relation of the Higgs condensate
value $\VEV v$ to the gravitational parameters is, up to
coefficients which may contain the above large logarithm, just the
geometric mean of the cosmological-constant and Planck/GUT scales
\begin{equation}
\VEV v^2 \sim M\mu \ . \label{eq:v}
\end{equation}

It is very unlikely that such simple-minded relations should
exist. But if the chance is nonvanishing, then it is extremely
attractive to pursue the idea further, because it is most likely
that the relations will only hold if the underlying dynamics at
and above the electroweak scale is austere and simple, instead of
being the very rich and complicated scenario more typically
presumed to be the case. The most viable option for simplicity is
something like the ``desert" scenario, with little if any new
physics this side of the short distance frontier at the GUT/Planck
scales. Note that the usual hierarchy-problem argument against
this scenario is inoperative. From the form of the above
expression, Eq.~(\ref{eq:v}), for the Higgs $\VEV v$, it is clear
that the explanation for its value would be as deep as the
explanation for the value of the cosmological constant, and
probably correlated.

I cannot resist going one step further at this point, although it
is a little off the main track of this subject matter. Recently
there has been a proposal that black hole interiors exist in a
different vacuum phase than their Schwarzschild exteriors, in
particular that they are nonsingular and described by a static
deSitter metric \cite{ref:h}. This means that the cosmological
constant in the interior of a black hole is different from its
exterior value, and therefore serves as an order parameter which
distinguishes the two phases. The magnitude of the cosmological
constant scales as the inverse square of the radius of the black
hole:
\begin{equation}
\mu^4 \sim \frac{M^2}{R^2} \ .
\label{eq:w}
\end{equation}
Consequently the Higgs $\VEV v$ and other Standard Model
parameters also are different in a black hole interior. For
example, in the interior of the black hole at the center of our
galaxy, the electroweak $\VEV v$ would increase by about 4 orders
of magnitude, while the ratio of electron to proton mass would
presumably diminish by about a factor 20, and the ratio of pion
mass to proton mass by a factor 5.  In general, within an
infinitely large black hole, the Standard Model would become
noninteracting and trivial, while for a Planck-scale black hole it
would become strongly interacting. This scenario has very
interesting implications for the evolutionary cosmology advocated
by Smolin \cite{ref:i}.

The implications of this picture for our model of the Goldstone
photon are also interesting, and that is the main reason for this
digression. We have chosen our notation with this in mind: the
parameters $M$ and $\mu$ in this section are (provisionally) to be
identified with those in the previous section. Especially
interesting is the form of the effective potential in Eq.
(\ref{eq:r}).  In the limit of vanishing cosmological constant it
also vanishes. This is quite consistent with the behavior of the
rest of the Standard Model dynamics which we have described.

A simple model which might place these ideas in more concrete
terms would be to assume our universe to be the (static DeSitter)
interior of a Reissner-Nordstrom black hole, with an approximately
critical amount of charge on the horizon, created somehow via
spontaneous breakdown. The electrostatic potential in the DeSitter
interior is then of the same order of magnitude $M$ as the
estimate inferred in the previous section. I have not yet worked
this out in any detail. But no matter how it turns out, I do not
think this is a very realistic model. Surface charge on a black
hole horizon is expected to be screened out. And the electroweak
gauge bosons, as well as the QCD gluons, deserve a similar
Goldstone treatment, and the details of their emergence will
certainly interact with those of the Goldstone photon, leading to
textures in spacetime and in momentum space of greater richness
and complexity.

\section{Lorentz Noncovariance}

The spontaneous Lorentz symmetry breaking probably leaves in its
wake residual noncovariant effects, along with possible violations
of gauge invariance. There clearly exists a preferred reference
frame in the formalism, naturally identified with the frame for
which the cosmic background radiation is locally at rest. Most
insidious is the violation of gauge invariance.  Our emergent QED
is defined in temporal gauge, and in that gauge the potentials are
in principle observable. This is hazardous, because there is no
energetic reason in the gauge-invariance approximation for gauge
potentials to be small.  Therefore small coefficients in front of
gauge-dependent terms in the Hamiltonian or Lagrangian may not be
sufficient to keep their physical effects small.

We have not attempted a thorough study of these effects, and here
only sketch out some of the issues. We recall that the QED
effective Lagrangian we arrived at looks completely standard
\begin{equation}
\mathcal{L} = \frac{1}{e^2}\, \left[\frac{1}{2}(E^2-B^2)-j_\mu
A^\nu\right] + \cdots
 \label{aaa}
\end{equation}
where irrelevant terms have been dropped, and where
\begin{equation}
\longvec E = - \dot{\longvec A} - \longvec\nabla\phi \qquad
\longvec B = \longvec\nabla \times \longvec A \qquad j_\mu =
\frac{e^2J_\mu}{G} \ .
 \label{bbb}
\end{equation}
However the definition of the electric field is non-standard
because the scalar potential has been eliminated in terms of a
nonlinear function of the vector potential $\longvec A$
\begin{equation}
\phi(\longvec A) = \sqrt{M^2 + \longvec A^2} - M = \frac{A^2}{2M}
+ \cdots \ .
 \label{ccc}
\end{equation}
As a consequence, the equations of motion contain extra terms,
with coefficients which are inverse powers of the GUT/Planck mass.
A short calculation gives for the equation of motion
\begin{equation}
\dot{\longvec E} = \longvec j + \left(\longvec\nabla \times
\longvec B\right) + \Gamma \longvec\beta \ .
 \label{ddd}
\end{equation}
In the above expression, the correction-term field $\beta$ is
defined as
\begin{equation}
\longvec\beta = \frac{\longvec A}{\sqrt{M^2+A^2}} \ .
 \label{eee}
\end{equation}
Note that its magnitude is bounded above by unity.  Note also that
the quantity
\begin{equation}
\Gamma = \longvec\nabla \cdot \longvec E - j_0
 \label{fff}
\end{equation}
which we may call the Gauss-law field, multiplies this new
correction term.

This equation of motion must account for two of the four Maxwell
equations, the other two source-free equations following
automatically from the definition of the fields.  The Gauss'-law
piece follows from taking the divergence of Eq.~(\ref{ddd}):
\begin{equation}
\frac{\partial}{\partial t}\, \longvec\nabla \cdot \longvec E =
\longvec \nabla \cdot \longvec j + \longvec\nabla \cdot
\left[\longvec\beta \Gamma\right] = \frac{\partial j_0}{\partial
t} + \longvec\nabla \cdot \left[\longvec\beta \Gamma\right] .
 \label{ggg}
\end{equation}
This is more conveniently written
\begin{equation}
\frac{\partial\Gamma}{\partial t} = \longvec\nabla \cdot
\left[\longvec\beta \Gamma\right] .
 \label{hhh}
\end{equation}
Without the correction term, this implies that the Gauss-law field
$\Gamma$ is time independent.  In canonically quantized QED, this
is implemented as a constraint equation on the Hilbert space. Only
states that are annihilated by the Gauss-law field are deemed
physically admissible.  With the correction term, this situation
still persists, albeit marginally.   If the Gauss-law field
$\Gamma$ vanishes at some initial time, then Eq.~(\ref{hhh}) shows
that this will be true at subsequent times.  The result is that,
given this constraint, all the Maxwell equations survive without
correction. The effects of the nonlinear correction terms are
limited to complicating the relationship of the gauge potentials
$\longvec A$ to field strengths $\longvec E$ and $\longvec B$.

This is encouraging, but hardly the end of the story.  Loop
corrections might change this situation.  And there seems to be an
inherent instability or at least metastability in the structure of
this effective action.  In addition, the generating functional
$W(J)$ obtained by integrating out all fields $A_\mu$ appears to
contain nonlinear terms, leading {\em e.g.} to small anomalous
$N$-body $(N\ge 3)$ Coulomb-like interactions between charged
sources. And even in the gauge invariant terms, possible
noncovariant corrections generated by loops must be carefully
examined. Especially important are the noncovariant corrections to
the $(E^2-B^2)$ term, which are sensitive to the ultraviolet and
which are experimentally bounded to better than one part in
$10^{31}$ \cite{ref:m}.

\section{Summary}

The Goldstone photon discussed above was motivated initially by
the Nambu-Jona-Lasinio (NJL) Goldstone pion. In the intervening
forty years a great deal of progress has been made in the strong
interactions, and our perspective on what the NJL model means has
matured. To this day it still plays a rather prominent, albeit
controversial, role in chiral QCD. The work of Holger Bech's
colleague Mitya Diakonov \cite{ref:xxx} and his collaborators
provides an especially sharp example of the role of the NJL
picture within QCD. In brief, starting from the basic QCD action
expressed in terms of quarks and gluons, they first integrate out
the gluons, assuming that the dominant configurations are
instanton-induced. This leaves behind an effective action
describing interacting quarks not very dissimilar to the NJL
interaction, with occurrence of chiral symmetry breaking via
instanton zero modes.  After introducing ($\pi$ and $\sigma$)
bosonic fields built from quark bilinears, the quarks and $\sigma$
field can also be integrated out, leaving behind a nonlinear
chiral effective action describing the residual pionic Goldstone
modes.

While the details of the procedure, such as the assumed dominance
of instanton configurations, may be debatable \cite{ref:j}, the
bottom line is still the emergent effective action describing the
Goldstone pions.  This effective action contains essentially an
infinite number of terms.  Only a handful are operators of
dimension less than or equal four, essentially a kinetic energy
term and a derivative-free effective potential. The remainder will
be of negligible importance at sufficiently low energy scales.
Some coefficients of those terms are determined by symmetry
considerations. but in general their sizes are determined by
dimensional analysis; the only scale in the problem (in the limit
of massless up and down quarks) is $\Lambda_{QCD}$ (perhaps
multiplied by $4\pi$).  To view this physics from low energy
upward is to see an effective theory of massless, almost
noninteracting pions, (trivially) accurate over the many orders of
magnitude of energy scales below the natural QCD energy scale
\cite{ref:k}. It of course becomes rapidly inoperative as that
scale is approached.

Now let us consider the analogous situation for the Standard Model
gauge theories. The Standard Model action is composed of a finite
number of terms, all of dimension less than or equal four. Higher
dimension terms are assumed to be strictly absent. If the gauge
bosons are indeed emergent, then one must expect that an infinite
number of terms in the action will be present, with a strength
scaled by the size of the ultraviolet cutoff $M$, which we have
taken to be of order GUT/Planck size. The terms of dimension
greater than four will be extremely small at present energy
scales.  In practice they will be not very relevant, provided they
do not break symmetries such as Lorentz covariance or gauge
invariance. However, this may not be the case for terms violating
gauge invariance, so special care should be taken in searching
phenomenologically for the possible presence of such terms
\cite{ref:m}.

Most of the above conclusions are not especially novel, because
any attempt, such as string theory, to incorporate gravity within
the Standard Model is likely to also generate an effective action
for the Standard-Model fields, described by a power series in the
gravitational constant.  However, what we described in the
previous sections for the Goldstone photon goes somewhat further.
First of all there is the aforementioned issue of noncovariance,
which may not need to be addressed in theories with increasingly
large symmetries at increasingly large energy scales, but almost
certainly must be addressed in theories with emergent degrees of
freedom (in the sense implied by analogy to condensed-matter
systems), where low-energy symmetries are increasingly broken as
the energy scale increases \cite{ref:b}.

In addition, the structure of terms in the action with dimension
less than or equal four is in our case different. The terms with
no derivatives of the gauge fields, which comprise an overall
effective potential, are in the Standard Model action forbidden to
be present. And this almost remains true in the picture we have
described. Unlike the case of the chiral effective action, there
is for these terms an assumed suppression factor of $\mu/M$, with
$\mu$ the cosmological constant scale and $M$ the GUT/Planck
scale. This suppression is not gratuitous, but introduced to
protect the phenomenology from gross Lorentz-noncovariant effects.
Even in the Higgs sector the suppression factor exists for the
Higgs mass term, assumed ({\em cf.} Eq. (\ref{eq:v})) to be of
order $\mu/M$. However, the quartic Higgs boson self-coupling is
for some reason not suppressed; this is a way of re-expressing the
hierarchy problem in this language. In addition, it has been
assumed that all terms involving spacetime derivatives of $A_\mu$
obey current conservation (other than the triangle anomaly).  In
particular this means that the dimension-four term $(\partial_\mu
A^\mu)^2$ is absent or strongly suppressed. In summary, we must in
general assume

\begin{enumerate}
\item
Fermion loop terms allowed in standard gauge theory are present,
with typically  a coefficient which is enhanced relative to the
standard radiative-correction contribution by a factor $\sim \ell
n(M/\mu)$.
\item
Terms disallowed in standard gauge theory, such as $A^2$, $A^4$,
$(\partial_\mu A^\mu)^2$, are suppressed by at least one power of
$\mu/M$.
\item
Mass terms in general (in particular the Higgs mass term) are
suppressed by at least one power of $\mu/M$.
\item
If for some reason there are intrinsically noncovariant
contributions, they too are suppressed by positive powers of
$\mu/M$.
\end{enumerate}
The actual power of $\mu/M$ which multiplies the leading
contribution to the effective potential need not be unity;
arguments can easily be found for a quadradic or even a quartic
dependence.  But of course it will be necessary to go further to
see whether any such pattern of suppressions can be motivated by
real physics.

At this point we have arrived at a situation very close to what
has been recently discussed by Holger Bech and his collaborators
\cite{ref:CFN}.  They argue that spontaneously broken Lorentz
covariance plus strict non-observability of noncovariant effects
requires the effective action of the spontaneously broken theory
to be identical to the gauge-invariant Standard-Model action. We
arrive at their results in the limit $\mu \rightarrow 0$, but
prefer to not quite go to the limit.  No matter what choice is
made, what is really needed are concrete examples of the mechanism
of spontaneous symmetry breakdown.

We also assumed that the Standard Model coupling strengths are
connected to the gravitational parameters. Here we note that
the behavior which was assumed can be established in a simple and
elegant way. One multiplies the entire Standard Model Lagrangian
density (expressed at the GUT/Planck scale) by a factor $\ell n\
M^2/\mu^2$, sets the interior coupling constants to values of
order unity, and then rescales all fields, fermionic as well as
bosonic, to create properly normalized kinetic energy terms. The
behavior of the coupling constants then follows Eq.~(\ref{eq:u}),
implying that as the cosmological constant $\mu^4$ tends to zero,
all the Standard Model interactions vanish.

\section{Outlook}

These considerations are not worth much unless the underlying
physics, if any, of the spontaneous breakdown is delineated. I
personally am attracted to the ideas of Volovik \cite{ref:zzz},
which of course rely on earlier work of Holger Bech in a major
way. Particularly attractive is the argument for a nearly
vanishing cosmological constant: if the vacuum is analogous to a
droplet of quantum liquid in equilibrium at very low temperature,
the pressure vanishes, up to boundary-surface corrections.
Vanishing vacuum pressure is the same statement as vanishing
cosmological constant. This simple argument in my mind creates a
new setting of the cosmological-constant problem: it becomes the
understanding of Standard Model particles as collective
excitations of this purported quantum liquid. The original problem
does not go away, but is restated in terms of why the collective
modes of such a quantum liquid so faithfully respect gauge
invariance and Lorentz covariance, as well as general covariance
for the emergent graviton. Very important in the Volovik viewpoint
is the structure and topology of momentum space. It may be there
that we may find the textures needed for understanding the
existence of not just the Goldstone photon but of all twelve
Goldstone gauge bosons. The condensed-matter phenomenon of ``Fermi
points" in momentum space, where for dynamical reasons the
fermionic excitations in their neighborhood take on Dirac-like or
Weyl-like character, may be an important analog \cite{ref:l}.
Perhaps there is an array of such Fermi points in the vacuum
momentum space that can account for the large number of fermion
species and related gauge excitations.

The problem of the graviton as an emergent degree of freedom must
also be addressed. It presumably must be faced at a more
fundamental level, {\em e.g.} higher energy scale, than the gauge
sector. There probably needs to be a condensate associated with it
as well. If one does not choose to invent new degrees of freedom,
about the only possibility seems to be to somehow exploit the
right-handed, gauge-singlet, Majorana neutrino condensates present
in the see-saw mechanism of neutrino mass generation.

All of this is a big order. The QCD example of the remote
relationship of fundamental theory to the chiral effective theory,
as well as a host of similar examples from condensed matter
theory, does not provide encouragement that one can comprehend the
underlying mechanisms, if any, of emergence. On the other hand,
the phenomenology of the Standard Model, including the behavior of
its parameters that we have assumed, is simpler than those
examples. There is a lot more symmetry. There are many excitations
but only one limiting velocity of propagation for them.  And the
dynamic range over which the effective theory is valid is huge.
The many Standard-Model parameters of mass and mixing must serve
as valuable, albeit enigmatic clues. A skeptic may argue that
these differences from historical examples indicate that the
emergence idea is inapplicable. But it just might be possible,
however unlikely, that the situation to be faced is in fact
simpler and more comprehensible than those examples. Given any
nonvanishing chance that this is true, this provides more than
enough motivation for an optimist to vigorously pursue this
program.

Preparation of this note has been greatly aided by conservations
with Marvin Weinstein and Michael Peskin.  I also thank my
black-hole colleagues Ron Adler, Pisin Chen, Robert Laughlin, and
David Santiago, for helpful conversations and encouragement.
Finally, it should be clear that this work owes a great deal to
the many creative and seminal ideas of Holger Bech Nielsen.

\end{document}